\documentclass[prb,twocolumn,showpacs,showkeys,amssymb]{revtex4}

\usepackage{graphicx}% Include figure files
\usepackage{dcolumn}% Align table columns on decimal point
\usepackage{bm}% bold math

\begin{document}

%\preprint{}
%\tightenlines

\title{Non-reciprocity and heat transfer in far from equilibrium processes}
\author{Alejandro Cabo Montes de Oca }
\affiliation{Instituto de Cibernetica, Matematica y Fisica, Calle E
 309, Vedado, Ciudad Habana, Cuba}
\email{cabo@cidet.icmf.inf.cu}

\begin{abstract}
\bigskip
A non-reciprocal phonon model for microwave or optical isolators
is considered. It gives a simpler framework to further investigate
the formerly argued possibility for a heat transfer between black
bodies at common temperatures. While the non-dissipative device
retains the Detailed Balance property, the presence of dissipation
breaks it. This property allows a net transfer of heat between the
two black bodies at common temperatures, whenever the absorptive
elements are at lower temperatures than the one being common to
the bodies. 

\end{abstract}


\maketitle

%%%%%%%%%%%%%%%%%%%%%%%%%%%%%%%%%%%%%%%%%%%%
%% MAINMATTER
%%%%%%%%%%%%%%%%%%%%%%%%%%%%%%%%%%%%%%%%%%%%

%\section{Introduction}
In this work we continue the previously started examination of the
physics of the heat transfer among bodies and the fact that such
energy flows are expected to always spontaneously occur from
higher to lower temperature regions. This property is equivalent
to the Law of increasing Entropy\cite{zeman}. However, in our view
its theoretical derivation from statistical physics deserves more
close attention. A concrete fact underlining this need is the lack
of definition of the entropy for physical processes occurring far
from the equilibrium situation\cite{ruelle}. It can be understood
that for a very wide class of physical systems having reciprocal
dynamics, the heat will always travel from higher to lower
temperature regions. However, when the bodies have a
non-reciprocal dynamical response, this property is no so
naturally evident. Non-reciprocal phenomena occurs in practice
within devices such as optical and microwave isolators and
\cite{lax,lax1,hand1, hand2,hand3}and more generally in all the
physics of the magnetism having so many applications in
technology.

The former paper \cite{cabo1} made an analysis of two types of
isolators connecting two regions containing black-body radiation
at common temperatures.  In the present article we 
consider the realization of the same kind of
phenomena within a simple soluble 1D non-reciprocal phonon model. It is
defined by a string having a uniform charge density which is
placed within a magnetic field orientated along the equilibrium
line of the string. The systems wave modes are explicitly
determined. They show two independent Faraday waves for each
direction of propagation. These are employed for defining a phonon
version of the optical and microwave isolators discussed in
\cite{cabo1}.

A main new possibility introduced here is the addition of a
lateral string (the ''vane'') at one of the ends of the isolator
which simulates the dissipative resistance vane in the microwave
or optical versions of these devices. It is then shown that when
the vane is disconnected, the transmission coefficients through
both sides is identical no matter the frequency or the
orientations of the slits simulating the optical polarizers. This
result in fact represent a clear solution of the so called Wien
paradox in optical system \cite{Bazarov}. Also it illustrates the
validity of the Detailed Balance principle for the two terminal
conservative version of the isolator. After that, the waves modes
for the system including the vane  were also solved. The calculated spectrum for the
transmission coefficients implied the breaking of the Detailed
Balance principle for the isolator including the vane. However, it
follows that when the three outputs are at the same temperature,
the net flux through any of the three terminals exactly vanish.
This result excludes the presence of the heat flux unbalance under
full thermal equilibrium. However,  a net heat
flow between Black-Bodies at common temperatures is predicted  when the vane is
placed in contact with a very low temperature heat reservoir.

%%\section{A non-reciprocal phonon model}
Seeking for a simple model for the optical and microwave isolators
discussed in \cite{cabo1} let us consider a Lagrangian system
formed by a uniformly charged string in presence of an also
uniform magnetic field. The Lagrangian of the system and its Euler
equations are given by
\begin{eqnarray}
L &=&\int dx(\frac{\rho _m}2(\partial _t\overrightarrow{u})^2-\frac{k_h}2%
(\partial _x\overrightarrow{u})^2+  \nonumber \\
&&\frac{\rho _q}{2c_s}\overrightarrow{u}.\overrightarrow{u}\times
\overrightarrow{B}), \\
0 &=&-\rho _m\partial _t^2\overrightarrow{u}+k_h\partial _x^2\overrightarrow{%
u}+\frac{\rho _q}{2c}(\partial _t\partial
_x\overrightarrow{u})\times \overrightarrow{B}.
\end{eqnarray}
\noindent where $\overrightarrow{u}(x,t)$ are the vector
displacement of the points of the string with respect to their
equilibrium positions and $\rho _m $ and $\rho _q $ are the linear
mass and charge densities, $k_h$ is the Hook law constant. The
magnetic field is oriented along the string in the x direction
$\overrightarrow{B}=(1,0,0)$. It should be mentioned that the self
interaction of the charged string will be disregarded in order to
simplify the discussion. The coulomb interaction, basically will
produce the appearance of a plasmon mode introducing a gap at low
frequencies. Searching for the oscillation modes in the usual form
\[
\overrightarrow{u}(x,t)=\exp (i.kx-ik_ox_o)\overrightarrow{u}(k)
\]
in which $k_o=w/c_s,x_o=c_st$ and $c_s=\sqrt{k_h/\rho _m}$ is the
sound velocity in the cord, c is the light velocity. The
Euler equation in (1) after written in components takes the form
\[
\left( (k^2-k_o^2){\large \delta }_{ij}+\frac{i\rho _qk_oc_s}{2k_hc}{\LARGE %
\epsilon }_{ilj}B_l\right) u_j=0,
\]
\noindent which after to be solved produce the dispersion
relations for longitudinal and transversal waves for positive
energy values $\epsilon =\hbar \,k_oc_s$ 
$
k_o = k, k_o = \sigma M+\sqrt{k^2+M^2},\sigma =\pm, 
M  = \frac{\mid \rho _qB\mid c_s}{2k_hc}.
$

\noindent Inversely, the momenta and frequencies obey the
relations
$
k_\sigma =\sqrt{(k_o-\sigma M)^2-M^2}$,  with $\sigma =\pm.  
$
It should be noticed that for a fixed energy $\epsilon =\hbar
\,k_oc_s$ there are two solutions having momenta $k_{+}$ and
$k_{-}$ . After performing linear combinations of two modes
associated with the same fixed positive frequency, Faraday waves
can be obtained. That is, they show the Faraday effect:  their
polarizations are linear at any point and its angular orientation
changes proportionally with the displacement along the string.

A phonon version of the experimentally existing microwave
or optical isolators investigated in \cite{cabo1} is
accomplished by considering an arrangement depicted in Fig. 1.

\begin{figure}[ht] 
\begin{center}
\includegraphics[width=.9\linewidth,angle=0]{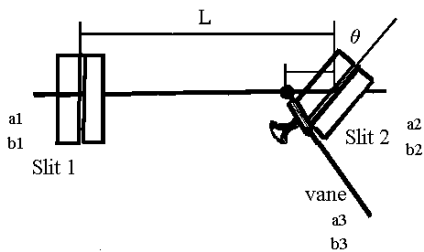} 
\caption{Phonon Isolator} 
\end{center} 
\end{figure} 
%\begin{figure}
% \includegraphics[height=.3\textheight]{isolator}
%  \caption{Phonon Isolator}
%\end{figure}
The device can be described in the following way: a) Two slits
forming an angle $\theta $ among them are situated a distance L
restricting the movement of a long piece of the string described
in previous section. They lay within a plane being orthogonal to
the string and it is assumed that there is no friction between
them and the cord. In other words they form conservative
mechanical constraints on the system. b) At a distance $\delta \ll
L$ of one of the them, let us say the right one, a second string
is attached to the main one, by mean of a bar sliding without
friction  inside a cylinder at rest. That is, the point at the
junction is constrained to move linearly in a direction orthogonal
to the initial string. This fact implies that the newly added
string, that we will call from now on, the ''vane'', is allowed to
oscillate only longitudinally. c) Let us also consider that the
frequency lies in the linear part of the dispersion relations,
that is, when $M\ll k_o$. Then, the length $L\;$is chosen to
assure that a Faraday mode propagating to the right and starting
at the slit 1 being polarized along it, after arriving to the slit
2 will be also polarized along it. This property, in turns will
assure that the waves incident on the left of slit 1, being
polarized along it will have a transmission coefficient equal to
one, whenever the coupling with the vane is disconnected. d) Next,
the angular orientation of the vane at the fixing point distant a
length $\delta $ from the slit 2 is selected to coincide with the
exact polarization of the before analyzed Faraday wave having
transmission coefficient equal to one. e) At last, let us consider
that outside the two slits, the charge density of the string
vanish and also that additional non-dissipative mechanical
constraints exist which only allow the transversal polarization of
the waves in the directions of the associated slits. The only
purpose of this assumption is to reduce the number of independent
wave modes of the system in order to simplify the discussion. It
should be notice that in the proposed arrangement the longitudinal
oscillations do not couple with the transversal ones (within the
linear approximation only).

The above conditions, then assure that the system can be described
by a three terminal scattering matrix \cite{Dicke}: $
b_i=S_{ij} \, a_j, \, i,j=1,2,3,$ where the pair $(a_1,b_1)$ contains the
incident $a_1$ and reflected $b_1$ wave amplitudes at the slit one and
$(a_2,b_2),(a_1,b_1)$ are the corresponding quantities at the slit
2 and the vane respectively.

\noindent In order to solve for the matrix S, auxiliary Faraday
modes propagating in the two internal regions, shown in Fig 1 were
defined. Then, the amplitudes for such modes were related among
themselves by imposing the continuity of the amplitudes and the
components of the forces along the directions of the slits 1 and 2
and the vane's direction. Further, after defining the scattering
modes associated to the input, let say the one with number
$i(i=1,2,3)$, as such wave modes for
which the incident amplitude $a_i=1$ and all the other incident amplitudes 
$a_j=0,\ j\neq i$ at the rest of the terminals vanish, the set of
linear equations defined by the boundary conditions were solved.
More precisely, as one of the propagating modes shows a gap for
low frequencies, this wave turns to be spatially damped boundary
mode when the frequency is low enough to be in the gap. The
effects of these low frequency oscillations were disregarded in
this initial discussion. 

The above described equations for finding the scattering modes
were solved initially for the case in which the vane is
disconnected from the central string. This situation corresponds
to an optical or microwave isolator in
which the ''resistive element'' absorbing the power in the output 2 is absent
\cite{lax}. The system is also completely analog to the one
related with the so called Wien paradox \cite{Bazarov}. The
transmission coefficient for the resulting two terminal device 
$ T =\frac{\mid b_2\mid }{\mid a_1\mid },
a_2 = 0, $ for the wave incident upon the slit one was then evaluated as a
function of the wave vector and the angle $\theta $ between the
two slits. The results are shown in the two dimensional plot in
Fig. 2. It, should be reminded that the transmission coefficient
for the wave coming in through the slit 2 can be extracted from
the figure by changing the sign of the angle $\theta.$ The data
correspond to a length $L\ $ for which the angle $\theta=\pi/4$
produces a value one for the transmission coefficient from left to
right at high frequencies.
\begin{figure}[ht] 
\includegraphics[width=3in]{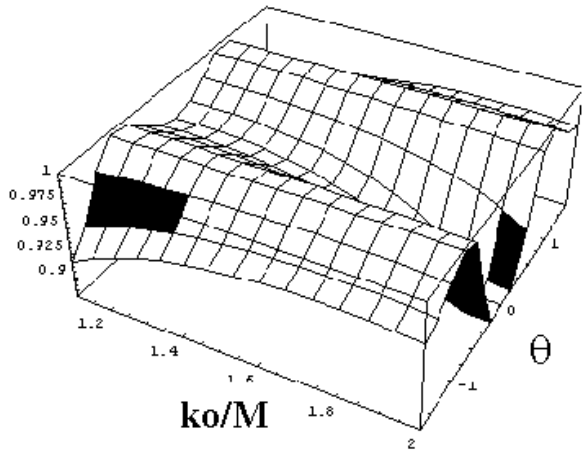} 
\caption{Transmission coefficients for the Phonon Isolator without losses}  
\end{figure} 
The picture remarkably shows that in the absence of the vane, in
spite of the non-reciprocity of the system, the transmission
coefficient associated to the two inputs are exactly equal
whatever the value of the angle and the frequency. In other words,
the fraction of the incident power transmitted from one side to
another is exactly the same whatever side is taken as the input.
The equality to one of the transmission for the angle $\theta
=\frac \pi 4$ is also evidenced. The results curiously indicate that
in spite of the multiple reflections between the two slits of the
wave entering through slit 2, the transmission coefficient for
this wave turns to be equal to  one. This conclusion gives a clear
solution of the above mentioned Wien paradox \cite{Bazarov} and
also implies the exact validity of the Detailed Balance principle
for the non-dissipative isolator. In conclusion, the absence of a
net heat flux between two black-bodies joined by such isolators
also follows.

Further, the case in which the vane is connected was investigated.
At this point it may be useful to remark that we considered in
detail the region of frequencies $M\ll k_o$ in which the rotation
of the polarization per unit length is a constant and its value
for the length $L$ was selected to produce the transmission
coefficient $T=1.$ Therefore, for this situation, we evaluated the
two transmission coefficient 
$t_{12}(k_o) = \frac{\mid b_1\mid ^2}{\mid a_2\mid ^2}, 
a_1 =a_3=0, $ and  $t_{13}(k_o) = \frac{\mid b_1\mid ^2}{\mid a_3\mid ^2},
a1 =a_2=0, $ in the now three terminal device and for the angle $\theta $
between the slits equal to $\frac \pi 4.$

As the reflection coefficient for the wave incident at the slit 1
is equal to zero, the question about a possible power flow
unbalance at the input number one is determined by these
quantities. The values of the transmission coefficients are
illustrated at Fig. 3 as functions of the energy in the range
$k_o/M\in (3,10)$ where the linearity condition of the dispersion
relations is obeyed very approximately.

Figure 3 clearly indicates a breaking of the detailed Balance in
the three terminal system simulating the optical and microwave
isolators. This conclusion follows from the fact that
$t_{12}(k_o)\neq t_{21}(k_o)=1$. The possibility for the existence
of physical systems not satisfying this property was elderly
advanced by the same Boltzmann\cite {Bazarov} long time ago. 

However, let us now  argue that under the
assumption of the thermal contact of the three terminals with heat
reservoirs characterized by  a common value of the temperature,
the net heat flux through any of the inputs is equal to zero. For
the slit number one this outcome is a direct consequence of the
results in Fig. 3 and the identical formula for the link between
the flux of energy and the amplitudes of the waves in each of the
outputs. This last property is determined by the assumed common
physical parameters (density $\rho _m$ and Hook law constant $k_h$
) of the strings associated to any of the terminals. The relevant
property of Fig. 2 determining the vanishing of the heat flux at
the left of the slit 1 is the relation
$t_{12}(k_o)+t_{13}(k_o)=1$, which is clearly satisfied by the
data in the picture. This is also the situation for any of the other terminals
as a general consequence of the conservative character of the three
terminal device (See Ref.\cite{Dicke}), which implies that the $S$
matrix is unitary. To see this explicitly, it is useful to
consider that the columns of $S$ defined by a fixed value of $j$,
give the components $b_i^{(j)},i=1,2,3,$ of the scattering modes
normalized by the condition $a_j=1$ and $a_l=0$ for $l\neq j $.
Therefore, let us consider the power $P_e^{(i)}$of the waves
emerging from output $i$ under the assumption that equal powers
are entering through the three terminals due to the common
temperature conditions. Then it follows $
P_e^{(i)} = p_o\sum_{j=1}^3b_i^{(j)}(b_i^{(j)})^{*}=p_o
\sum_{j=1}^3S_{ij}\ S_{ij}^{*}= p_o a_ia_i^{*}=p_o $
where $p_o$ is the constant relating the square of the amplitudes
(defined by some convention) with the associated power flow in a
travelling wave. The above relation implies that the power of the
outgoing waves at any of the terminals is identical to the power
of the thermal incident waves entering through the same input,
whenever all the heat reservoirs are at a common temperature $T$.
\begin{figure}[ht] 
\includegraphics[width=3in]{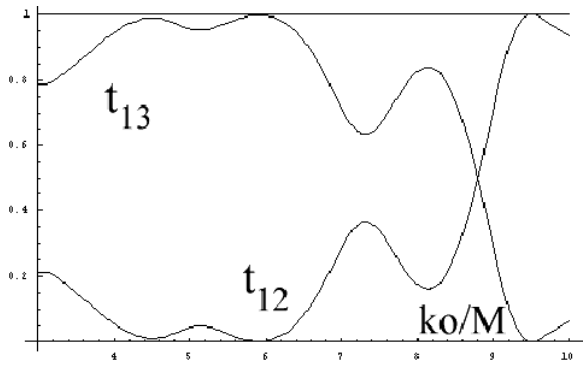} 
\caption{Transmission coefficients determining the outgoing
heat flow in the input at the left of the slit 1}  
\end{figure} 
%\begin{figure}
%  \includegraphics[height=.3\textheight]{break3}
%  \caption{Transmission coefficients determining the outgoing
%  heat flow in the input at the left of the slit 1}
%\end{figure}
%\section{Heat flow between Black-Bodies at common temperatures}

Finally,  we will consider that the vane terminal is laying
at a very much small temperature $T_v<<T$ where $T$ is the
temperature of the left and right heat reservoirs in contact with
the pieces of the central string arriving at the left and right
slits 1 and 2. Also the angle $\theta $ will be selected to
furnish a transmission coefficient equal to one for the waves
incident through the slit 1. Under this conditions it follows that
at least for a very extended region of the energies for which the
relation $M\ll k_o$ is valid, the net flux passing from left to
right of the isolator is greater than the one circulating in the
opposite sense. Since this property is independent of the
temperature, we will assume that it can be elevated sufficiently
enough for to be able for disregarding the low energy region, when
deciding about the question of the net heat balance. Under this
central assumption, it can be concluded from the data in Fig. 2
that the system will produce a heat flow passing from the Black
-Body at the left of the slit 1 to the black body at the right of
slit 2.  This conclusions motivate to consider in further works the possibility
for a reduction of the statistical entropy through concrete mechanisms for the
extraction of energy through the vane. 
 In ending, I would like to acknowledge the discussions with
Dr. A. Gonzalez, C. Trallero and F. Comas and the general support of 
Dr. D. Sheehan, the Org. Committee of the Conference and the
Associate Scheme of the  ASICTP.
%%%%%%%%%%%%%%%%%%%%%%%%%%%%%%%%%%%%%%%%%%%%%%%%
%% BACKMATTER
%%%%%%%%%%%%%%%%%%%%%%%%%%%%%%%%%%%%%%%%%%%%%%%%
%\begin{theacknowledgments}
%I would like to acknowledge the helpful discussions with Dr. A.
%Gonzalez, C. Trallero and F. Comas.
%\end{theacknowledgments}
%% If the bibliography is produced without BibTeX comment out the
%%\section{REFERENCES}
%%\bibliographystyle{aipproc}   % if natbib is available
%%\begin{references}

%%\end{references}
\bibliography{sample}

\IfFileExists{\jobname.bbl}{}
 {\typeout{}
  \typeout{******************************************}
  \typeout{** Please run "bibtex \jobname" to optain}
  \typeout{** the bibliography and then re-run LaTeX}
  \typeout{** twice to fix the references!}
  \typeout{******************************************}
  \typeout{}
 }

\end{document}